\documentclass[10pt,conference,letterpaper]{IEEEtran}
\usepackage{cite}
\usepackage{amsmath,amssymb,amsfonts}
\usepackage{algorithm}
\usepackage{algorithmic}
\usepackage{graphicx}
\usepackage{textcomp}
\usepackage{xcolor}
\usepackage{bbm}
\usepackage{booktabs}
\usepackage{nccmath}
\usepackage{url}
\usepackage{subfig}
\usepackage[capitalise]{cleveref}
\usepackage{glossaries}
\usepackage{pgfplots}
\pgfplotsset{compat=1.16}
\newcommand{\E}{\mathbb{E}}

\title{Remote Anomaly Detection in Industry 4.0 Using Resource-Constrained Devices}

\author{\IEEEauthorblockN{Anders E. Kal{\o}r, Daniel Michelsanti, Federico Chiariotti, Zheng-Hua Tan and Petar Popovski}
\IEEEauthorblockA{Department of Electronic Systems\\
Aalborg University, Denmark\\
Email: \{aek,danmi,fchi,zt,petarp\}@es.aau.dk}}

\begin{document}

\maketitle

\begin{abstract}
A central use case for the Internet of Things (IoT) is the adoption of sensors to monitor physical processes, such as the environment and industrial manufacturing processes,  where they provide data for predictive maintenance, anomaly detection, or similar. The sensor devices are typically resource-constrained in terms of computation and power, and need to rely on cloud or edge computing for data processing. However, the capacity of the wireless link and their power constraints limit the amount of data that can be transmitted to the cloud. While this is not problematic for the monitoring of slowly varying processes such as temperature, it is more problematic for complex signals such as those captured by vibration and acoustic sensors. In this paper, we consider the specific problem of remote anomaly detection based on signals that fall into the latter category over wireless channels with resource-constrained sensors. We study the impact of source coding on the detection accuracy with both an anomaly detector based on Principal Component Analysis (PCA) and one based on an autoencoder. We show that the coded transmission is beneficial when the signal-to-noise ratio (SNR) of the channel is low, while uncoded transmission performs best in the high SNR regime.
\end{abstract}

\begin{IEEEkeywords}
Remote monitoring, anomaly detection, source coding, channel coding
\end{IEEEkeywords}

\section{Introduction}
Over the past few years, the Internet of Things (IoT) has connected objects, sensors, and appliances of all kinds. One of the most promising future applications is the Industrial IoT~\cite{varga20205g}, i.e., the use of IoT technology and standards to help automate and monitor industrial processes. In particular, remotely monitoring a process by using cheap, low-power wireless sensors can significantly reduce operation costs of machinery in remote or dangerous environments, such as windmills or photovoltaic farms~\cite{bedi2018review}, flagging anomalies in the sensory data and requesting human intervention when needed.

The strict size, energy, and complexity requirements that make IoT solutions attractive for these applications pose significant challenges. The standard techniques used in Industry 4.0 contexts are often complicated and rely on an abundance of data, often using deep learning over massive datasets~\cite{zhou2020variational}. Naturally, this does not fit an IoT scenario, in which the sensors are resource-constrained in terms of power and computation capabilities. Furthermore, the devices cannot transmit complete measurements at a high rate for processing in the cloud, such as raw sound or vibration recordings, as they would quickly deplete their batteries, as well as congesting the radio channel. Instead, a more reasonable approach is to let the devices transmit \emph{feature vectors} of their measurements.

However, the statistics of anomalous conditions are often hard to model and may not be known a priori~\cite{ukil2016iot}, which makes appropriate source coding complex: optimal quantization requires knowledge of the input distribution, and anomalies outside this range are likely to suffer from high distortion. On the other hand, applying universal quantization techniques increases the distortion for nominal samples, which also makes it difficult to discriminate between nominal and anomalous samples. In general, quantizers tend to suppress anomalies, and as a result, uncoded transmission can be a valid alternative in some cases~\cite{gastpar2003code}. In fact, analog transmission might even outperform digital encoding in terms of sample reconstruction accuracy in the short blocklength regime~\cite{kostina2013lossy}. This has been observed in other practical deep learning tasks over channels, such as classification~\cite{jankowski2020joint} and image retrieval~\cite{jankowski2020wireless}.

In this work, we consider the problem of remote anomaly detection in an edge-computing scenario, in which a wireless sensor has constrained computational and communication resources, and transmits feature vectors to an edge server that performs the anomaly detection. We compare uncoded and coded transmission of the feature vectors, and in line with the typical anomaly detection situation, we assume that only data from nominal conditions are known, and that anomalies follow unknown statistics outside the nominal operational region. We consider two types of anomaly detectors at the edge node. The first is based on Principal Component Analysis (PCA)~\cite{lakhina2004diagnosing}, and while being analytically and computationally tractable, it can only detect linearly separable anomalies. The second detector is an autoencoder, which is based on a neural network and can detect more complex anomalies. We first examine the PCA method on a constructed scenario to reveal the fundamental differences between the two transmission schemes, and then evaluate the autoencoder on a dataset of real acoustic data from industrial equipment. Our results show that for both the PCA method and the autoencoder, coded transmission is best when the channel signal-to-noise ratio (SNR) is low, while uncoded transmission performs best in the high SNR regime.

The rest of the paper is divided as follows. The system model is presented in Sec.~\ref{sec:system}, along with the considered coded and uncoded transmission schemes. PCA anomaly detection is then described and analyzed in Sec.~\ref{sec:pca}, while an autoencoder-based anomaly detector is introduced in Sec.~\ref{sec:autoenc}.
The numerical results for both the PCA and autoencoder methods are shown in Sec.~\ref{sec:results}, and finally Sec.~\ref{sec:concl} concludes the paper.

\section{System model}\label{sec:system}
We consider an edge-computing scenario comprising a single device, which has a wireless connection to a base station equipped with an edge server. While the device is resource-constrained and can only do a limited computation, the edge server is assumed to be unconstrained. Based on a sensor observation, the device constructs a sample $\mathbf{x}\in\mathbb{R}^N$, also referred to as a feature vector, which either belongs to the class of nominal observations, or to the alternative class of anomalous observations. Without loss of generality, we shall assume that the sample vector is normalized, i.e. $(1/N)\E[\|\mathbf{x}\|^2]=1$ where $\|\cdot\|$ is the $\ell_2$ norm. Due to the resource constraints of the device, the classification is assumed to take place at the edge server, and thus the device needs to transmit the sample before it can be classified.

In line with existing literature on anomaly detection, we will assume that the statistics are known only for nominal samples and not anomalies. Such statistics are typically gathered from a dataset assumed to include only nominal samples. Furthermore, we will restrict our focus to residual based anomaly detectors, where $\mathbf{x}$ is determined based on the \emph{reconstruction error} $\|\hat{\mathbf{x}}-\mathbf{x}\|$, where $\hat{\mathbf{x}}=g^{-1}\left(g(\mathbf{x})\right)$ is the reconstruction of $\mathbf{x}$ after applying some transformation. The intrinsic assumption is that by properly choosing $g(\mathbf{x})$, the residual for a nominal sample is small but large for an anomalous sample. In particular, $\mathbf{x}$ is declared an outlier if the magnitude of the residual exceeds a given threshold $\delta$
\begin{equation}
  \|\hat{\mathbf{x}}-\mathbf{x}\|^2 > \delta^2,
\end{equation}
and otherwise the sample is assumed to be nominal, see e.g., ~\cite{jackson1979control,lakhina2004diagnosing,purohit2019mimii}. Note that $\delta$ controls the trade-off between sensitivity and the false positive probability.
To characterize the performance of a detector, we define the accuracy for a dataset with $P$ positive and $N$ negative samples as
\begin{equation}
  \text{acc} = \sup_{\delta}\, \frac{P\Pr(\text{TP}|\delta)+N(1-\Pr(\text{FP}|\delta))}{P+N},
\end{equation}
where $\Pr(\text{TP}|\delta)$ and $\Pr(\text{FP}|\delta)$ are the true positive and false positive probabilities, respectively.

In this initial paper, we restrict ourselves to considering the additive White Gaussian Noise (AWGN) channel, which is instructive to compare the coded and uncoded transmission strategies. Prior to transmission, the device encodes the sample to $D$ symbols using the encoder $f:\mathbb{R}^N\to\mathbb{R}^D$. Throughout the paper we will assume that $D\ge N$. Denoting the encoded sample by $\bar{\mathbf{x}}=f(\mathbf{x})$, which is assumed to be normalized with respect to the symbol power, $(1/D)\E[\|\bar{\mathbf{x}}\|^2]=1$, the signal received at the receiver is
\begin{equation}
  \mathbf{y} = \bar{\mathbf{x}} + \mathbf{z},
\end{equation}
where $\mathbf{z}\sim\mathcal{N}(\mathbf{0}, \Gamma^{-1}\mathbf{I})$ is additive white Gaussian noise. The signal-to-noise ratio (SNR) is then equal to $\Gamma$. We shall denote the decoded feature vector by $\mathbf{x}'=f^{-1}(\mathbf{y})$, and note that this channel can be seen as a stationary instance of a block-fading channel where the noise power is determined by the amount of fading. We consider both coded and uncoded transmission strategies as outlined next.

\subsection{Coded transmission}
With the coded transmission strategy, the device first quantizes its sample with a given rate (i.e. to a given number of bits), and then transmits the quantized sample over the channel. We assume that the device uses a capacity-achieving channel code so that the transmission rate is $C=(1/2)\log_2(1+\Gamma)$ measured in bits per (real) symbol. We note that this reflects an idealized scenario and thus is an upper bound on what can be achieved in practice where channel estimation, etc. is needed. Under this assumption, the total number of bits that the device has available to represent the sample is $B=CD$ bits.

Because of resource constraints, we assume that the device employs scalar quantizers for each entry in the sample vector designed to minimize the expected squared error distortion of the nominal samples (since only these are known). The quantizers are designed in two steps. First, the $B$ bits are allocated to the individual entries in the sample based on their expected distortion by iteratively allocating one bit to the entry with the highest expected distortion (see e.g.~\cite{gershogray} for a discussion on this greedy approach). For simplicity, we assume that the entries in the sample vector are Gaussian so that the expected distortion of the $i$-th entry is given by $\sigma_i^2 2^{-2b_i}$, where $\sigma_i^2$ is the variance of the entry and $b_i$ is the number of allocated bits. After this procedure we have $\sum_{i=1}^N b_i=B$. In the second step, the individual quantizers for each sample entry are designed using Lloyd's algorithm~\cite{lloyd82}, such that $2^{b_i}$ quantization points are allocated to entry $i$. To encode a sample vector, the device simply picks, for each entry $i$ in the vector, the quantization point from the set of $2^{b_i}$ points that is closest to entry $i$ in the sample vector.

\subsection{Uncoded transmission}
In the uncoded transmission scheme, the device transmits the sample vector using uncoded analog (real) symbols. Since there is no need for source and channel coding, this scheme has the advantage of simplifying the transmitter device. Note that the analog symbols can be concatenated with a coded fragment, e.g. a header to allow for metadata, device addressing, etc., but this part is not important for this paper. To match the number of available symbols $D$, the transmitter spreads the signal using a matrix $\mathbf{Q}\in\mathbb{R}^{D\times N}$ satisfying  $\mathbf{Q}^T\mathbf{Q}=\mathbf{I}$, i.e. the encoder and decoder are $f(\mathbf{x})=\mathbf{Q}\mathbf{x}$ and $f^{-1}(\mathbf{y})=\mathbf{Q}^T\mathbf{y}$, respectively. When the sample is transmitted uncoded, the received signal can be written as
\begin{equation}
  \mathbf{x}' = \mathbf{x} + \mathbf{w},
\end{equation}
where $\mathbf{w}\sim\mathcal{N}\left(\mathbf{0},\frac{N\Gamma^{-1}}{D}\mathbf{I}\right)$.

\section{PCA based anomaly detection}\label{sec:pca}
In this section, we study a PCA based anomaly detector to gain insight into the impact of the channel on the detection accuracy. We first introduce the PCA subspace method, and then quantify the false positive and true positive probabilities for a given threshold $\delta$. Due to the difficulties associated with modelling the quantization effects for low rates (which are our main interest), we limit our analysis to the case of uncoded transmissions, and note that this also includes the case without a channel as the special case with $\Gamma^{-1}=0$. The performance of the coded transmission scheme with quantization effects will be evaluated using Monte Carlo simulations. Although the PCA subspace method is general, to ease exposition and to keep the analysis tractable, we will assume that the nominal samples follow a Gaussian distribution with zero mean, i.e. $\mathbf{x}\sim\mathcal{N}(\mathbf{0}, \mathbf{\Sigma})$ where $\text{tr}(\mathbf{\Sigma})=N$ to satisfy the normalization assumption. We will denote the eigenvalues of $\mathbf{\Sigma}$ corresponding to the normalized eigenvectors by $\hat{\sigma}_1\ge\hat{\sigma}_2\ge\ldots\ge\hat{\sigma}_N$.

\subsection{PCA subspace method}
The PCA subspace method has been applied successfully to many practical problems, see e.g.~\cite{lakhina2004diagnosing}, and while more sophisticated non-linear methods for anomaly detection exist, such as kernel methods and deep autoencoders~\cite{hoffmann07,sakurada14}, these methods are typically difficult to analyze and share many similarities with the PCA method.

We assume that the received (decoded) samples $\mathbf{x}'$ have zero mean and covariance matrix $\mathbf{\Sigma}'$, but in general the distribution does not have to be Gaussian. This assumption is valid both for the coded and uncoded policies---in particular, in the uncoded case $\mathbf{\Sigma}'=\mathbf{\Sigma}+(N/D)\Gamma^{-1}\mathbf{I}$. Let $\mathbf{V}\mathbf{\Lambda}\mathbf{V}^T=\mathbf{\Sigma}'$ be the eigendecomposition of $\mathbf{\Sigma}'$ with normalized eigenvectors ordered by the eigenvalues $\lambda_1 \ge \lambda_2\ge\ldots\ge \lambda_N$, and denote by $\hat{\mathbf{V}}\in\mathbb{R}^{N\times K}$ the matrix composed by the $K$ eigenvectors of $\mathbf{\Sigma}$ with the largest eigenvalues. The PCA subspace method decomposes a sample $\mathbf{x}'$ (which is either nominal or anomalous) into two components
\begin{equation}
    \mathbf{x}'=\hat{\mathbf{x}}+\tilde{\mathbf{x}},
\end{equation}
where $\hat{\mathbf{x}}$ and $\tilde{\mathbf{x}}$ are referred to as the \emph{modeled} and \emph{residual} components, respectively.
The modeled component is the projection of $\mathbf{x}'$ onto the subspace spanned by $\hat{\mathbf{V}}$, i.e. $\hat{\mathbf{x}}=\hat{\mathbf{V}}\hat{\mathbf{V}}^T\mathbf{x}'$, and $\tilde{\mathbf{x}}=(\mathbf{I}-\hat{\mathbf{V}}\hat{\mathbf{V}}^T)\mathbf{x}'=\tilde{\mathbf{P}}\mathbf{x}'$. As with other residual based anomaly detectors, the assumption is that, provided that a sufficient number of eigenvectors are included in $\hat{\mathbf{V}}$, the residual vector $\tilde{\mathbf{x}}$ is going to be small in magnitude when $\mathbf{x}'$ follows the assumed model, but large when $\mathbf{x}'$ is anomalous.

\subsection{False positive probability}
False positives occur when a nominal sample $\mathbf{x}_0'$ exceeds the detection threshold $\delta$. We are interested in characterizing
\begin{equation}
  \Pr(\text{FP}|\delta)=\Pr(\|\tilde{\mathbf{x}}_0'\| > \delta).
\end{equation}
As mentioned, we restrict our analysis to the uncoded case where $\mathbf{x}_0'\sim\mathcal{N}(\mathbf{0}, \mathbf{\Sigma}')$ and $\mathbf{\Sigma}'=\mathbf{\Sigma}+(N/D)\Gamma^{-1}\mathbf{I}$. In this case, the residual vector $\tilde{\mathbf{x}}_0'=\tilde{\mathbf{P}}\mathbf{x}_0'$ is also Gaussian with zero mean and covariance matrix $\tilde{\mathbf{\Sigma}}'=\tilde{\mathbf{P}}\mathbf{\Sigma}'\tilde{\mathbf{P}}^T$. It can be shown that $\|\tilde{\mathbf{x}}_0'\|^2$ is distributed as
\begin{equation}
    \|\tilde{\mathbf{x}}_0'\|^2\sim\sum_{i=K+1}^N\lambda_i Z_i^2,\label{eq:normdist_fpr}
\end{equation}
where $\lambda_i$ is the $i$-th largest eigenvalue of $\mathbf{\Sigma}'$ and $Z_i$ are i.i.d. standard Gaussian random variables~\cite{jackson1979control}.
To make the impact of the additive noise more explicit, the expression can also be written using the eigenvalues of $\mathbf{\Sigma}$ by noticing that $\lambda_i=\hat{\sigma}_i+(N/D)\Gamma^{-1}$ where $\hat{\sigma}_i$ is the $i$-th largest eigenvalue of $\mathbf{\Sigma}$.
The distribution of the squared magnitude of the residual $\|\tilde{\mathbf{x}}_0'\|^2$ can then be approximated by the quantity
\begin{equation}
    c(\|\tilde{\mathbf{x}}_0'\|^2)=\frac{\theta_1\left(\left(\|\tilde{\mathbf{x}}_0'\|^2/\theta_1\right)^{h_0} - 1 - \theta_2h_0(h_0-1)/\theta_1^2\right)}{\sqrt{2\theta_2h_0^2}},
\end{equation}
which follows a standard Gaussian distribution where $\theta_j=\sum_{i=K+1}^N\left(\hat{\sigma}_i+(N/D)\Gamma^{-1}\right)^j$ and $h_0=1-2\theta_1\theta_3/(3\theta_2^2)$~\cite{jackson1979control}. Thus, to compute the false positive rate, we can evaluate
\begin{equation}
  \Pr(\|\tilde{\mathbf{x}}_0'\| > \delta) \approx 1-\Phi(c(\delta^2)),\label{eq:magnitude_normal_approx}
\end{equation}
where $\Phi(\cdot)$ is the cumulative distribution function of a standard Gaussian distribution.

\subsection{True positive probability}
To derive the true positive probability, let us assume that an anomaly is reflected as an additive term in the sample $\mathbf{x}_{\mathbf{f}}=\sqrt{\eta}\mathbf{x}_0+\mathbf{f}$, where $\mathbf{x}_0$ represents a sample under nominal conditions and $\eta=1-(1/N)\|\mathbf{f}\|^2$ is introduced to have $(1/N)\E[\|\mathbf{x}_{\mathbf{f}}\|^2]=1$ (assuming $\|\mathbf{f}\|^2\le N$).
After being transmitted over the noisy channel, the received sample $\mathbf{x}_{\mathbf{f}}'=\mathbf{x}_{\mathbf{f}}+\mathbf{w}$ is Gaussian with mean $\mathbf{f}$ and covariance $\mathbf{\Sigma}'=\eta\mathbf{\Sigma}+(N/D)\Gamma^{-1}\mathbf{I}$.
The residual vector of the anomalous sample can be written
\begin{align}
    \tilde{\mathbf{x}}_{\mathbf{f}}' &= \tilde{\mathbf{P}}\mathbf{x}_{\mathbf{f}}'=\tilde{\mathbf{x}}_0'+\tilde{\mathbf{f}},
\end{align}
where $\tilde{\mathbf{x}}_0'=\tilde{\mathbf{P}}(\sqrt{\eta}\mathbf{x}_0+\mathbf{w})$ and $\tilde{\mathbf{f}}=\tilde{\mathbf{P}}\mathbf{f}$. We define the true positive probability as
\begin{equation}
  \Pr(\text{TP}|\delta)=\Pr(\|\tilde{\mathbf{x}}_{\mathbf{f}}'\| > \delta).
\end{equation}
In order for an anomaly $\mathbf{f}$ to be detectable by the PCA method, it must leave a nonzero residual, i.e. $\tilde{\mathbf{f}}\neq \mathbf{0}$ ($\mathbf{f}$ must not be in the null space of $\tilde{\mathbf{P}}$). Furthermore, $\mathbf{f}$ must leave a sufficiently large residual so that $\|\tilde{\mathbf{x}}_{\mathbf{f}}'\|>\delta$. Due to these conditions, the false positive probability generally depends on the anomaly vector distribution. In this work, we will restrict the analysis to the conditional false positive probability given an anomaly vector $\mathbf{f}$, i.e. $\Pr(\|\tilde{\mathbf{x}}_{\mathbf{f}}'\|>\delta|\mathbf{f})$. The marginal false positive probability can be obtained by averaging the conditional distribution over any distribution of the anomaly vectors.

Because we condition on $\mathbf{f}$, and thus $\tilde{\mathbf{f}}$ is deterministic, the residual vector is distributed as $\tilde{\mathbf{x}}_{\mathbf{f}}\sim\mathcal{N}(\tilde{\mathbf{f}}, \tilde{\mathbf{\Sigma}}')$ where  $\tilde{\mathbf{\Sigma}}'=\tilde{\mathbf{P}}\mathbf{\Sigma}'\tilde{\mathbf{P}}^T$. By generalizing the result from~\cite{jackson1979control} to the case with non-zero mean we obtain
\begin{equation}
    \|\tilde{\mathbf{x}}_{\mathbf{f}}\|^2\sim\sum_{i=K+1}^N\lambda_i (Z_i + t_i)^2,
\end{equation}
where $t_i$ is the $i$-th entry of $\mathbf{t} = \mathbf{V}^T(\eta\mathbf{\Sigma}+(N/D)\Gamma^{-1})^{-1/2}\tilde{\mathbf{f}}$, see e.g.~\cite{mathai1992quadratic}. This distribution can be approximated like before as
\begin{equation}
  \Pr(\|\tilde{\mathbf{x}}_{\mathbf{f}}'\| > \delta|\mathbf{f}) \approx  1-\Phi(c(\delta^2)),
\end{equation}
where $\theta_j$ are computed as~\cite{jensen1972gaussian}
\begin{equation}
  \theta_j=\sum_{i=K+1}^N \lambda_i^j(1+jt_i^2).
\end{equation}
Here we may again exploit that $\lambda_i=\left(\eta\hat{\sigma}_i+(N/D)\Gamma^{-1}\right)$ to write the expression in terms of the eigenvalues of $\mathbf{\Sigma}$.

\section{Autoencoder based anomaly detection}\label{sec:autoenc}
The main limitation of the PCA based method is that it requires anomalous samples to be linearly separable from the nominal ones. A more sophisticated alternative is to use an autoencoder, which is an artificial neural network that learns a compact representation of the input. Specifically, an autoencoder is typically composed of a set of encoder layers and a set of decoder layers, separated by a low-dimensional bottleneck layer, see e.g.~\cite{sakurada14}. The network is then trained to minimize the reconstruction loss between the input and the output given as
\begin{equation}
  J(\phi) = \lVert \mathbf{x}' - h(\mathbf{x}'|\phi) \rVert^2,
\end{equation}
where $\mathbf{x}'$ is the decoded input feature vector at the receiver, $h(\mathbf{x}'| \phi)$ denotes the output of the autoencoder and $\phi$ indicates the autoencoder parameters. Similar to the PCA method, the reconstruction error of the autoencoder can be used for anomaly detection by assuming that anomalous samples produce large reconstruction errors. However, contrary to the PCA the autoencoder is not limited to detecting linear anomalies in the samples. This flexibility comes at the cost of being more difficult to analyze, and thus we only consider a numerical evaluation of the autoencoder based method.

\section{Numerical results}\label{sec:results}
\subsection{PCA results}
We first evaluate the two transmission schemes for the PCA method on a constructed scenario that matches the assumptions of the analysis. Specifically, we assume that nominal samples are Gaussian with zero mean and covariance matrix $\mathbf{\Sigma}$. Because the PCA is independent to orthonormal transformations of the eigenvectors, the results only depend on the eigenvalues of $\mathbf{\Sigma}$ under the constraints that $\mathbf{\Sigma}$ has full rank and $\text{tr}(\mathbf{\Sigma})=\sum_{n=1}^N\hat{\sigma}_n=N$. We consider case where the eigenvalues of $\mathbf{\Sigma}$ are given as $\hat{\sigma}_1=50\beta,\hat{\sigma}_2=40\beta,\hat{\sigma}_3=30\beta,\hat{\sigma}_4=20\beta,\hat{\sigma}_5= 10\beta,\hat{\sigma}_5=\ldots=\hat{\sigma}_N=\beta$, where $\beta=1/(50+40+30+20+10+N-5)$ is to ensure that the eigenvalues sum to $N$. We study the case with $N=128$, and because the choice of eigenvectors represents a scenario with five dominant principal components we pick $K=5$. The fault vectors $\mathbf{f}$ are drawn uniformly from the $N$-sphere of various radii $\|\mathbf{f}\|^2$.

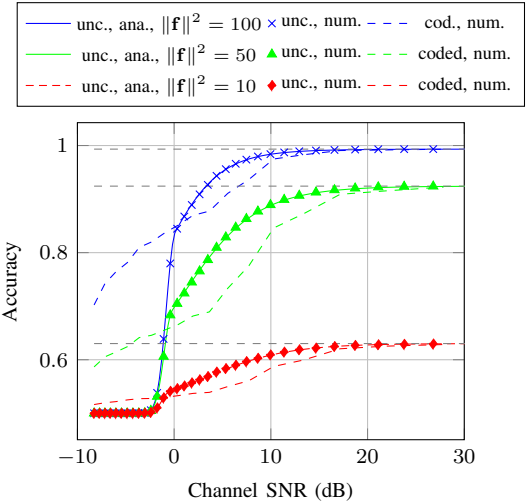
\begin{figure}
  \centering
  \begin{tikzpicture}
  \begin{axis}[
    xlabel={Channel SNR (dB)},
    ylabel={Accuracy},
    grid=both,
    font=\footnotesize,
    legend columns=3,
    legend style={
    font=\scriptsize,
    at={(0.5,1.05)},
    anchor=south,
    },
    xmin=-10, xmax=30,
    height=5.8cm,
    ]
    \addplot+[mark=none,blue] table[x index=0, y index=7] {pca_res_uncoded_ana.dat};
    \addlegendentry{unc., ana., $\|\mathbf{f}\|^2=100$};
    \addplot+[mark=x,only marks,blue] table[x index=0, y index=7,each nth point=3,filter discard warning=false, unbounded coords=discard] {pca_res_uncoded.dat};
    \addlegendentry{unc., num.}
    \addplot+[mark=none,blue,dashed] table[x index=0, y index=3,each nth point=3,filter discard warning=false, unbounded coords=discard] {pca_res_coded.dat};
    \addlegendentry{cod., num.}

    \addplot+[mark=none,green] table[x index=0, y index=5] {pca_res_uncoded_ana.dat};
    \addlegendentry{unc., ana., $\|\mathbf{f}\|^2=50$};
    \addplot+[mark=triangle*,only marks,green,mark options={fill=green}] table[x index=0, y index=5,each nth point=3,filter discard warning=false, unbounded coords=discard] {pca_res_uncoded.dat};
    \addlegendentry{unc., num.}
    \addplot+[mark=none,dashed,green,mark options={fill=green}] table[x index=0, y index=2,each nth point=3,filter discard warning=false, unbounded coords=discard] {pca_res_coded.dat};
    \addlegendentry{coded, num.}

    \addplot+[mark=none,red] table[x index=0, y index=3] {pca_res_uncoded_ana.dat};
    \addlegendentry{unc., ana., $\|\mathbf{f}\|^2=10$};
    \addplot+[mark=diamond*,only marks,red,mark options={fill=red}] table[x index=0, y index=3,each nth point=3,filter discard warning=false, unbounded coords=discard] {pca_res_uncoded.dat};
    \addlegendentry{unc., num.}
    \addplot+[mark=none,dashed,red] table[x index=0, y index=1,each nth point=3,filter discard warning=false, unbounded coords=discard] {pca_res_coded.dat};
    \addlegendentry{coded, num.}

    \addplot+[mark=none,dashed,gray] table[y index=8] {pca_res_uncoded_ana.dat};
    \addplot+[mark=none,dashed,gray] table[y index=6] {pca_res_uncoded_ana.dat};
    \addplot+[mark=none,dashed,gray] table[y index=4] {pca_res_uncoded_ana.dat};

  \end{axis}
\end{tikzpicture}
  \caption{Accuracy of the PCA subspace method with digital and analog transmission. The dashed horizontal lines indicate the channel-free accuracy.}
  \label{fig:pca_res}
\end{figure}

The results are shown in \cref{fig:pca_res}. As can be seen, the accuracy is highest for the coded transmission at low SNRs, while the uncoded transmission scheme provides higher accuracy once the SNR exceeds a given threshold around 0 dB. This indicates that for low SNRs, the additive noise from the uncoded scheme is dominant and makes it impossible to detect the faults. On the other hand, despite only transmitting few coarse features, the fact that the features only are affected by quantization noise makes the accuracy of the uncoded scheme higher at low SNRs.

\subsection{Autoencoder results}

To study the impact of feature coding in the more complex autoencoder setting, we consider the baseline autoencoder model proposed for the MIMII dataset in~\cite{purohit2019mimii}.
The MIMII dataset contains a collection of nominal and anomalous sounds of valves, pumps, fans, and slide rails from a factory environments. In particular, the sounds were recorded as 16-bit audio signals at 16 kHz with an eight-channel circular microphone array placed 50 cm away from the target machine (10 cm for valves). Besides the machine sounds, background noise signals from real factories were recorded with the same equipment and mixed with the recordings of the machine at three SNRs ($-6$ dB, $0$ dB, and $6$ dB).

In this work, we conduct experiments using only recordings from a single microphone. The recordings are used to construct feature vectors, which are the concatenation of five consecutive log-mel spectrogram frames, as further outlined in~\cite{purohit2019mimii}. As with other reconstruction based methods, the anomaly detection is performed by thresholding the reconstruction error. The autoencoder consists of six fully-connected layers of size 64, 64, 8, 64, 64, and 320, with a 320-element input vector. Each layer is followed by a ReLU activation function except for the last one which has linear activations.

For each SNR there are 26092 nominal and 6065 anomalous audio segments of 10 s, where the anomalous conditions include contamination, clogging, damage, etc. The anomalous segments are used only for testing along with a random subset nominal samples of the same size so that the testing dataset contains the same number of nominal and anomalous examples.
In line with our overall assumption, the baseline method used to detect the anomalous machines in~\cite{purohit2019mimii} is based on an unsupervised learning approach, where the training phase only includes the signals of nominal machines.

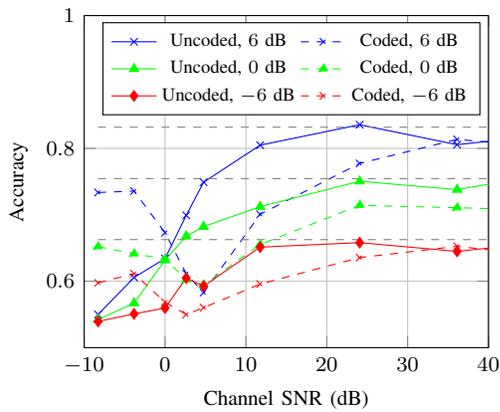
\begin{figure}
  \centering
  \begin{tikzpicture}
  \begin{axis}[
    xlabel={Channel SNR (dB)},
    ylabel={Accuracy},
    grid=both,
    font=\footnotesize,
    legend style={font=\scriptsize},
    legend columns=2,
    xmin=-10, xmax=40,
    ymin=0.5, ymax=1.0,
    height=6.0cm,
    ]
    \addplot+[blue,mark=x,mark options={fill=blue}] table[x index=0, y index=5] {autoencoder_res_uncoded.dat};
    \addlegendentry{Uncoded, $6~\text{dB}$};
    \addplot+[blue,mark=x,mark options={fill=blue},dashed] table[x index=0, y index=3] {autoencoder_res_coded.dat};
    \addlegendentry{Coded, $6~\text{dB}$};

    \addplot+[green,mark=triangle*,mark options={fill=green}] table[x index=0, y index=3] {autoencoder_res_uncoded.dat};
    \addlegendentry{Uncoded, $0~\text{dB}$};
    \addplot+[green,mark=triangle*,mark options={fill=green},dashed] table[x index=0, y index=2] {autoencoder_res_coded.dat};
    \addlegendentry{Coded, $0~\text{dB}$};

    \addplot+[red,mark=diamond*,mark options={fill=red}] table[x index=0, y index=1] {autoencoder_res_uncoded.dat};
    \addlegendentry{Uncoded, $-6~\text{dB}$};
    \addplot+[red,mark=x,mark options={fill=red},dashed] table[x index=0, y index=1] {autoencoder_res_coded.dat};
    \addlegendentry{Coded, $-6~\text{dB}$};
    
    \addplot+[dashed,gray,mark=none] table[y index=6] {autoencoder_res_uncoded.dat};
    \addplot+[dashed,gray,mark=none] table[y index=4] {autoencoder_res_uncoded.dat};
    \addplot+[dashed,gray,mark=none] table[y index=2] {autoencoder_res_uncoded.dat};
  \end{axis}
\end{tikzpicture}
  \caption{Accuracy of the autoencoder on the MIMII dataset for various channel SNRs. The number in the legend indicates the SNR of the sound recordings.}
  \label{fig:autoenc}
\end{figure}

The accuracy of the autoencoder anomaly detector with features transmitted over an AWGN channel is shown in \cref{fig:autoenc}. The autoencoder is trained for the specific SNR with data that includes noise (for the uncoded case) or is quantized (in case of coded transmissions). Furthermore, in the coded case, the quantizers are designed based on statistics estimated from the training set of nominal samples. As in the case of PCA, the coded scheme gives higher accuracy than the uncoded scheme when the SNR is low. However, somewhat surprisingly, the accuracy decreases as the SNR increases up to a certain point around 5 dB, before it starts to increase again. A possible reason for this may be that at low SNRs, the autoencoder is able to learn a good model of the few (coarse) features that are transmitted, while it is unable to learn a good model at slightly higher SNRs, where it receives many coarse features. In the high SNR regime, all features are transmitted with high resolution, and thus the autoencoder can again learn a good representation of the features. However, as was the case with PCA, the uncoded scheme generally performs better than the coded scheme in the high SNR regime.

\section{Conclusion}\label{sec:concl}
Motivated by wireless monitoring of processes in an industrial scenario using resource-constrained IoT devices, we have compared uncoded and coded transimssion schemes for anomaly detection. We have considered anomaly detection methods based on PCA and autoencoder, and both cases reveal that coded transmissions perform better at low SNRs while uncoded transmissions are better at high SNRs. Part of the reason is that at low SNRs, the noise from the uncoded transmission tends to be dominant and hide the anomalous parts, while the coded scheme provides coarse but less noisy samples. On the other hand, at high SNRs the quantizer in the coded transmission scheme, which is designed for nominal samples, is likely to suppress anomalous signals, whereas the anomalous signals in the uncoded scheme are only affected by the Gaussian noise.



\begin{thebibliography}{10}
\providecommand{\url}[1]{#1}
\csname url@samestyle\endcsname
\providecommand{\newblock}{\relax}
\providecommand{\bibinfo}[2]{#2}
\providecommand{\BIBentrySTDinterwordspacing}{\spaceskip=0pt\relax}
\providecommand{\BIBentryALTinterwordstretchfactor}{4}
\providecommand{\BIBentryALTinterwordspacing}{\spaceskip=\fontdimen2\font plus
\BIBentryALTinterwordstretchfactor\fontdimen3\font minus
  \fontdimen4\font\relax}
\providecommand{\BIBforeignlanguage}[2]{{%
\expandafter\ifx\csname l@#1\endcsname\relax
\typeout{** WARNING: IEEEtran.bst: No hyphenation pattern has been}%
\typeout{** loaded for the language `#1'. Using the pattern for}%
\typeout{** the default language instead.}%
\else
\language=\csname l@#1\endcsname
\fi
#2}}
\providecommand{\BIBdecl}{\relax}
\BIBdecl

\bibitem{varga20205g}
P.~Varga, J.~Peto, A.~Franko, D.~Balla, D.~Haja, F.~Janky, G.~Soos, D.~Ficzere,
  M.~Maliosz, and L.~Toka, ``{5G} support for industrial {IoT}
  applications--challenges, solutions, and research gaps,'' \emph{Sensors},
  vol.~20, no.~3, p. 828, 2020.

\bibitem{bedi2018review}
G.~Bedi, G.~K. Venayagamoorthy, R.~Singh, R.~R. Brooks, and K.-C. Wang,
  ``Review of internet of things (iot) in electric power and energy systems,''
  \emph{IEEE Internet of Things Journal}, vol.~5, no.~2, pp. 847--870, 2018.

\bibitem{zhou2020variational}
X.~Zhou, Y.~Hu, W.~Liang, J.~Ma, and Q.~Jin, ``Variational {LSTM} enhanced
  anomaly detection for industrial big data,'' \emph{IEEE Transactions on
  Industrial Informatics}, vol.~17, no.~5, pp. 3469--3477, 2020.

\bibitem{ukil2016iot}
A.~Ukil, S.~Bandyoapdhyay, C.~Puri, and A.~Pal, ``{IoT} healthcare analytics:
  The importance of anomaly detection,'' in \emph{30th International Conference
  on Advanced Information Networking and Applications (AINA)}.\hskip 1em plus
  0.5em minus 0.4em\relax IEEE, 2016, pp. 994--997.

\bibitem{gastpar2003code}
M.~Gastpar, B.~Rimoldi, and M.~Vetterli, ``To code, or not to code: Lossy
  source-channel communication revisited,'' \emph{IEEE Transactions on
  Information Theory}, vol.~49, no.~5, pp. 1147--1158, 2003.

\bibitem{kostina2013lossy}
V.~Kostina and S.~Verd{\'u}, ``Lossy joint source-channel coding in the finite
  blocklength regime,'' \emph{IEEE Transactions on Information Theory},
  vol.~59, no.~5, pp. 2545--2575, 2013.

\bibitem{jankowski2020joint}
M.~Jankowski, D.~G{\"u}nd{\"u}z, and K.~Mikolajczyk, ``Joint device-edge
  inference over wireless links with pruning,'' in \emph{2020 IEEE 21st
  International Workshop on Signal Processing Advances in Wireless
  Communications (SPAWC)}.\hskip 1em plus 0.5em minus 0.4em\relax IEEE, 2020,
  pp. 1--5.

\bibitem{jankowski2020wireless}
------, ``Wireless image retrieval at the edge,'' \emph{IEEE Journal on
  Selected Areas in Communications}, vol.~39, no.~1, pp. 89--100, 2020.

\bibitem{lakhina2004diagnosing}
A.~Lakhina, M.~Crovella, and C.~Diot, ``Diagnosing network-wide traffic
  anomalies,'' \emph{ACM SIGCOMM computer communication review}, vol.~34,
  no.~4, pp. 219--230, 2004.

\bibitem{jackson1979control}
J.~E. Jackson and G.~S. Mudholkar, ``Control procedures for residuals
  associated with principal component analysis,'' \emph{Technometrics},
  vol.~21, no.~3, pp. 341--349, 1979.

\bibitem{purohit2019mimii}
H.~Purohit, R.~Tanabe, K.~Ichige, T.~Endo, Y.~Nikaido, K.~Suefusa, and
  Y.~Kawaguchi, ``Mimii dataset: Sound dataset for malfunctioning industrial
  machine investigation and inspection,'' in \emph{Proceedings of the 4th
  Workshop on Detection and Classification of Acoustic Scenes and Events
  (DCASE)}, 2019.

\bibitem{gershogray}
A.~Gersho and R.~M. Gray, \emph{Vector quantization and signal
  compression}.\hskip 1em plus 0.5em minus 0.4em\relax Springer Science \&
  Business Media, 2012, vol. 159.

\bibitem{lloyd82}
S.~Lloyd, ``Least squares quantization in pcm,'' \emph{{IEEE} transactions on
  information theory}, vol.~28, no.~2, pp. 129--137, 1982.

\bibitem{hoffmann07}
H.~Hoffmann, ``Kernel pca for novelty detection,'' \emph{Pattern recognition},
  vol.~40, no.~3, pp. 863--874, 2007.

\bibitem{sakurada14}
M.~Sakurada and T.~Yairi, ``Anomaly detection using autoencoders with nonlinear
  dimensionality reduction,'' in \emph{Proceedings of the MLSDA 2014 2nd
  Workshop on Machine Learning for Sensory Data Analysis}, 2014, pp. 4--11.

\bibitem{mathai1992quadratic}
A.~M. Mathai and S.~B. Provost, \emph{Quadratic forms in random variables:
  theory and applications}.\hskip 1em plus 0.5em minus 0.4em\relax Dekker,
  1992.

\bibitem{jensen1972gaussian}
D.~R. Jensen and H.~Solomon, ``A gaussian approximation to the distribution of
  a definite quadratic form,'' \emph{Journal of the American Statistical
  Association}, vol.~67, no. 340, pp. 898--902, 1972.

\end{thebibliography}
\end{document}